\documentclass[conference]{IEEEtran}
\IEEEoverridecommandlockouts
\usepackage[utf8]{inputenc}
\usepackage[T1]{fontenc}
\usepackage{cite}
\usepackage{amsmath,amssymb,amsfonts}
\usepackage{algorithmic}
\usepackage{graphicx}
\usepackage{textcomp}
\usepackage{xcolor}
\usepackage{bm}
\usepackage{algorithm}
\usepackage{array}
\usepackage{url}
\usepackage{booktabs}
\usepackage{multirow}
\usepackage{makecell}
\usepackage[hidelinks]{hyperref}
\usepackage{balance}

\renewcommand{\IEEEkeywordsname}{Keywords}

\def\BibTeX{{\rm B\kern-.05em{\sc i\kern-.025em b}\kern-.08em
    T\kern-.1667em\lower.7ex\hbox{E}\kern-.125emX}}
\graphicspath{{./}}

\begin{document}

\title{Integrated Lander-Propulsion-GNC Framework for Autonomous Lunar Powered Descent}

\author{%
    \IEEEauthorblockN{Emre Aklan}
    \IEEEauthorblockA{Turkuzaysan Inc.\\
    Istanbul, Türkiye\\
    emre.aklan@turkuzaysan.com}
\and
    \IEEEauthorblockN{Fatih Seker}
    \IEEEauthorblockA{Turkuzaysan Inc.\\
    Istanbul, Türkiye\\
    fatih.seker@turkuzaysan.com}
\and
    \IEEEauthorblockN{Bekir Gencalioglu}
    \IEEEauthorblockA{Turkuzaysan Inc.\\
    Istanbul, Türkiye\\
    bekir.gencalioglu@turkuzaysan.com}
\and
    \IEEEauthorblockN{Mehmet Batuhan Kaya}
    \IEEEauthorblockA{Puura Inc.\\
    Istanbul, Türkiye\\
    mbatuhankaya98@gmail.com}
\and
    \IEEEauthorblockN{Yigit Serceoglu}
    \IEEEauthorblockA{Turkuzaysan Inc.\\
    Istanbul, Türkiye\\
    yigit.serceoglu@turkuzaysan.com}
\and
    \IEEEauthorblockN{Furkan Yavuz}
    \IEEEauthorblockA{Puura Inc.\\
    Istanbul, Türkiye\\
    furkan.yavuz@std.bogazici.edu.tr}
\and
    \IEEEauthorblockN{Omer Burak Iskender}
    \IEEEauthorblockA{Final Proximity Space Systems Inc.\\
    Istanbul, Türkiye\\
    burak@finalproximity.com\\
    \href{https://orcid.org/0009-0002-6143-8113}{ORCID: 0009-0002-6143-8113}}
\and
    \IEEEauthorblockN{\hspace{5.5cm}Burak Yaglioglu}
    \IEEEauthorblockA{\hspace{5.5cm}TUBITAK Space Technologies Research Institute\\
    \hspace{5.5cm}Ankara, Türkiye\\
    \hspace{5.5cm}burak.yaglioglu@tubitak.gov.tr\\
    \hspace{5.5cm}\href{https://orcid.org/0000-0001-5753-2831}{ORCID: 0000-0001-5753-2831}}
}

\maketitle
\vspace*{-\baselineskip}

\begin{abstract}
This paper presents an integrated lander-propulsion-GNC framework for autonomous lunar powered descent. The BUG VTVL test vehicle serves as the reference platform, with the YUNT V0 throttleable bipropellant engine providing variable thrust across a wide operating envelope, integrated with a real-time successive convexification guidance solver. The vehicle design accounts for structural configuration, landing stability, center-of-mass migration, and inertia evolution, while the propulsion architecture defines the throttle ratio, dead-zone behavior, and gimbal authority that constrain the guidance problem. A successive convexification algorithm addresses all nonconvexities; thrust lower bounds, mass depletion coupling, and thruster dead-zone behavior are all handled within a unified second-order cone program solvable in near-real time. Parametric analysis reveals a fundamental coupling between throttle ratio, pointing authority, and surface gravity. Monte Carlo simulations validate guidance robustness, achieving sub-50-meter landing precision under realistic perturbations.
\end{abstract}

\vspace{\baselineskip}

\begin{IEEEkeywords}
\textit{powered descent guidance, successive convexification, convex optimization, lunar landing, guidance and control.}
\end{IEEEkeywords}

\section{Introduction}
\label{sec:intro}

Autonomous powered descent and landing on the Moon presents extreme demands on fuel efficiency, actuator precision, and real-time trajectory planning. Recent landing failures, including SpaceIL's Beresheet (2019), ispace's Hakuto-R M1 (2023), and Russia's Luna 25 (2023), underscore the necessity for robust, integrated guidance and propulsion systems. These missions highlight the critical importance of tight integration between guidance algorithms and propulsion system characteristics, a gap this paper addresses.

Convex optimization has transformed powered descent guidance. Açikmeşe and Ploen~\cite{b_acik2007} demonstrated that the minimum-fuel problem admits an exact second-order cone program (SOCP) reformulation through lossless convexification, rigorously extended in~\cite{b_acik2011}. Blackmore et al.~\cite{b_blackmore2010} incorporated glide-slope and obstacle avoidance constraints, subsequently validated in JPL's G-FOLD flight experiments~\cite{b_gfold}. For nonlinear dynamics, successive convexification (SCvx)~\cite{b_mao2016, b_szmuk2020} iterates convex subproblems with trust regions, achieving convergence within 5 to 15 iterations and millisecond-scale solve times enabling real-time onboard replanning; Malyuta et al.~\cite{b_malyuta2022} provide a comprehensive tutorial. For full 6-DoF problems, dual quaternion representations~\cite{b_iskender_dq} provide a unified translational-rotational framework, and convex optimization has proven effective in safety-critical proximity operations~\cite{b_iskender_phd}. Constraint tightening approaches~\cite{b_iskender_ecc} can systematically shrink safe corridors to enhance landing accuracy under uncertainty.

Despite these advances, three critical gaps persist: (1)~systematic handling of all nonconvexities within a unified SOCP framework, (2)~practical treatment of thruster dead-zone behavior (a non-ideality often absent from standard formulations), and (3)~decoupled vehicle-guidance design approaches that neglect how mass property evolution during descent affects guidance feasibility and fuel consumption. This paper addresses these gaps through an integrated framework instantiated on the BUG VTVL test vehicle and YUNT V0 bipropellant engine, platforms intended for autonomous lunar soft landing. The framework jointly optimizes lander structure (Section~\ref{sec:design}), propulsion constraints (Section~\ref{sec:propulsion}), and convex guidance synthesis (Section~\ref{sec:gnc}), with comprehensive validation via parametric design-space exploration and Monte Carlo simulation under realistic lunar descent conditions (Section~\ref{sec:results}).

\section{Lander Design}
\label{sec:design}

\subsection{Structural Configuration and Design Rationale}

The BUG VTVL test lander (Turkuzaysan) employs a four-legged open-frame truss structure with a central body housing the main engine, propellant tanks, and avionics, following heritage NASA lunar lander configurations~\cite{b_nasa_design, b_wu_structural, b_weixiong_prelim}. Each landing gear consists of a primary strut and two secondary struts for load absorption~\cite{b_wang_legged}. The truss architecture minimizes structural mass while allowing flexible configuration tailoring.

\subsection{Landing Stability Analysis}

Landing stability depends critically on the footprint diameter and center-of-gravity height~\cite{b_witte_touchdown, b_sahinoz_landing, b_matsuura_overturn}. Table~\ref{tab:tipover_footprint} shows how tip-over angle improves with footprint diameter at fixed CoG height ($h = 1.0$~m).

\begin{table}[t]
\centering
\renewcommand{\arraystretch}{1.4}
\caption{Tip-Over Angle vs.\ Variable Footprint Diameter at Fixed CoG Height (h = 1.0~m)}
\label{tab:tipover_footprint}
\begin{tabular}{|c|c|c|}
\hline
\textbf{\makecell[c]{Footprint \\ Diameter (m)}} & \textbf{\makecell[c]{Distance to \\ Tipping Axis (m)}} & \textbf{\makecell[c]{Tip-Over \\ Angle (deg)}} \\
\hline
1.5 & 0.75 & 36.87 \\
\hline
2.0 & 1.00 & 45.00 \\
\hline
2.5 & 1.25 & 51.34 \\
\hline
3.0 & 1.50 & 56.31 \\
\hline
3.5 & 1.75 & 60.25 \\
\hline
\end{tabular}
\end{table}

Table~\ref{tab:tipover_cgheight} quantifies the effect of CoG height on tip-over angle at fixed footprint diameter ($D = 2.0$~m).

\begin{table}[t]
\centering
\renewcommand{\arraystretch}{1.4}
\caption{Tip-Over Angle vs.\ Variable CoG Height at Fixed Footprint Diameter (D = 2.0~m)}
\label{tab:tipover_cgheight}
\begin{tabular}{|c|c|c|}
\hline
\textbf{\makecell[c]{CoG Height \\ (m)}} & \textbf{\makecell[c]{Distance to \\ Tipping Axis (m)}} & \textbf{\makecell[c]{Tip-Over \\ Angle (deg)}} \\
\hline
0.50 & 1.0 & 63.43 \\
\hline
0.75 & 1.0 & 53.13 \\
\hline
1.00 & 1.0 & 45.00 \\
\hline
1.25 & 1.0 & 38.66 \\
\hline
1.50 & 1.0 & 33.69 \\
\hline
\end{tabular}
\end{table}

A key design trade-off emerges: increasing the footprint diameter enhances stability but adds structural mass, while lowering the center of gravity improves stability and reduces mass but may compromise ground clearance. The BUG reference vehicle achieves a tip-over angle of $\arctan\!\bigl(\tfrac{D/2}{h_{\text{CoG}}}\bigr) = \arctan(1.885/1.47) \approx 52^\circ$ with its wide stance, providing ample margin above the $29^\circ$ maximum guidance tilt angle.

\subsection{Reference Lander Configuration}

The BUG lander employs an inverted tripod landing gear configuration that minimizes design loads by creating nearly pure axial forces. Table~\ref{tab:bug_summary} summarizes key parameters: wet mass $m_0 = 259.5$~kg, dry mass $174$~kg, propellant mass fraction 33\%, and thrust-to-weight ratio $2.0$.

\begin{table}[t]
\centering
\renewcommand{\arraystretch}{1.4}
\caption{BUG Vehicle Parameters}
\label{tab:bug_summary}
\begin{tabular}{|c|c|}
\hline
\textbf{Parameter} & \textbf{Value} \\
\hline
Height $H$ & 2.26 m \\
\hline
Footprint diameter $D$ & 3.77 m \\
\hline
Aspect ratio $H/D$ & 0.60 \\
\hline
Wet mass $m_0$ & 259.5 kg \\
\hline
Dry mass $m_{\text{dry}}$ & 174 kg \\
\hline
Propellant mass & 85.5 kg \\
\hline
Thrust-to-weight ratio & 2.0 \\
\hline
\end{tabular}
\end{table}

As propellant is consumed, the center of mass migrates downward and the inertia tensor evolves. Table~\ref{tab:cg_inertia} summarizes the CoM location and principal moments of inertia at three propellant states. The 9~cm vertical CoM descent and 10\% reduction in transverse inertia are captured by the GNC model. Because propellant depletion lowers the center of gravity, landing stability at touchdown exceeds the initial wet-mass condition.

\begin{table}[t]
\centering
\renewcommand{\arraystretch}{1.4}
\caption{BUG CoM Migration and Principal Inertia (kg m$^2$)}
\label{tab:cg_inertia}
\begin{tabular}{|c|c|c|c|}
\hline
\textbf{Property} & \textbf{Wet} & \textbf{Mid (20 s)} & \textbf{Dry} \\
\hline
CoG$_y$ (m) & 1.47 & 1.45 & 1.38 \\
\hline
CoG$_z$ (m) & 0.01 & 0.00 & -0.01 \\
\hline
$J_{xx}$ & 140.2 & 136.7 & 126.1 \\
\hline
$J_{yy}$ & 96.3 & 93.6 & 86.5 \\
\hline
$J_{zz}$ & 140.1 & 136.5 & 126.0 \\
\hline
\end{tabular}
\end{table}

\begin{figure}[t]
\centerline{\includegraphics[width=\columnwidth]{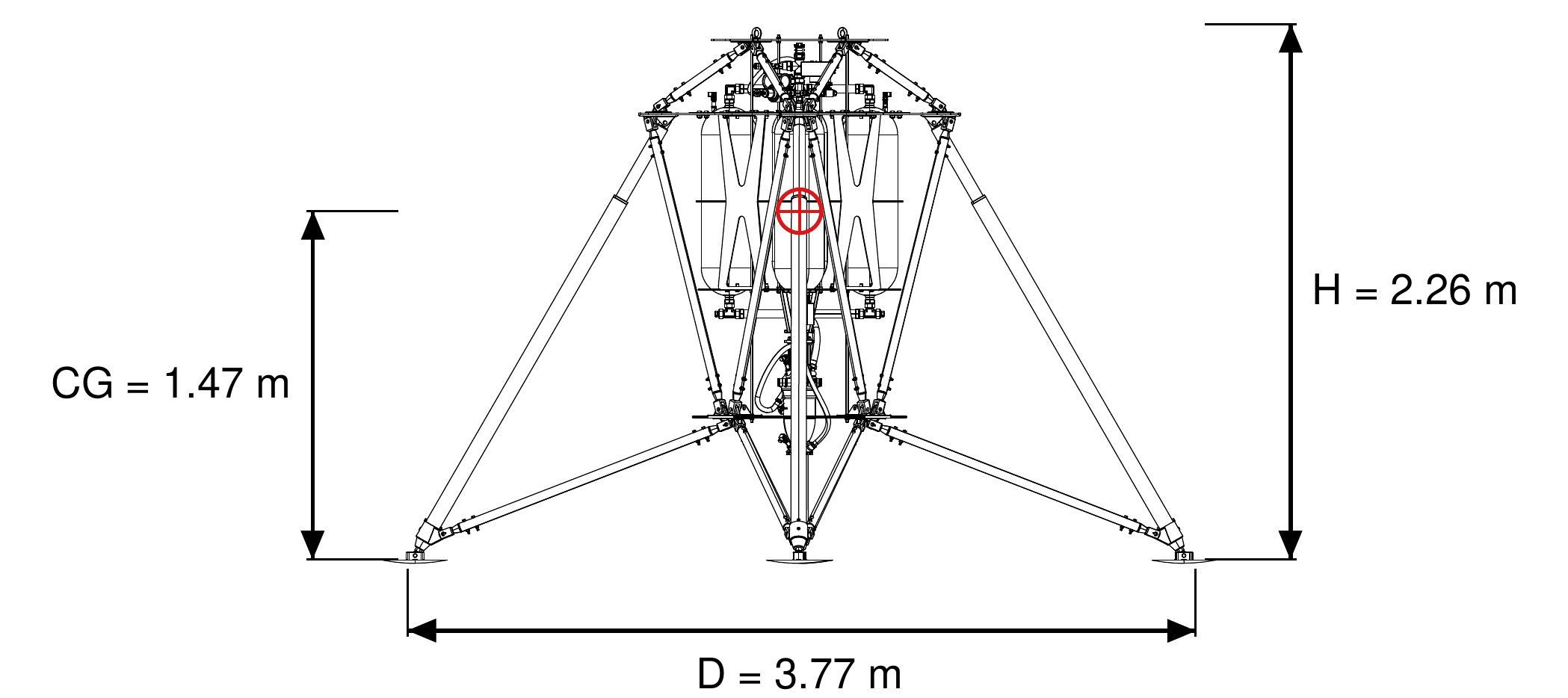}}
\caption{BUG landing test vehicle ($H=2.26$~m, $D=3.77$~m, $h_{\mathrm{CG}}=1.47$~m). Red $\oplus$ denotes centre of mass.}
\label{fig:lander_vehicle}
\end{figure}

Fig.~\ref{fig:lander_vehicle} shows the BUG test vehicle with a four-legged open-frame truss structure.

Table~\ref{tab:comparison} places BUG in context with representative VTVL test platforms (Colibri~\cite{b_colibri}, Xodiac~\cite{b_xodiac}, Morpheus~\cite{b_morpheus}) and recent lunar landers (Blue Ghost~\cite{b_blueghost}, Resilience~\cite{b_resilience}, IM-2~\cite{b_im2_clps}). BUG occupies a low aspect-ratio, wide-footprint regime tailored for stability margin during test flights.

\begin{table*}[t]
\centering
\renewcommand{\arraystretch}{1.4}
\caption{Comparison of BUG with VTVL Test Platforms and Lunar Landers}
\label{tab:comparison}
\small
\begin{tabular}{|c|c|c|c|c|c|c|c|}
\hline
& \multicolumn{4}{|c|}{\textit{VTVL Test Platforms}} & \multicolumn{3}{|c|}{\textit{Lunar Landers}} \\
\cline{2-8}
& \textbf{BUG} & \textbf{Colibri}~\cite{b_colibri} & \textbf{Xodiac}~\cite{b_xodiac} & \textbf{Morpheus}~\cite{b_morpheus} & \textbf{Blue Ghost}~\cite{b_blueghost} & \textbf{Resilience}~\cite{b_resilience} & \textbf{IM-2}~\cite{b_im2_clps} \\
\hline
Engine thrust (kN) & 0.8-5.2 & 2.2 & 3.2 & 20 & 1.0/1.6 & 0.4 & 3.1 \\
\hline
Wet mass (kg) & 260 & 200 & 300 & 1{,}050 & N/A & N/A & N/A \\
\hline
Height (m) & 2.26 & 2.5 & 3.5 & 3.0 & 2.0 & 2.3 & 4.0 \\
\hline
Footprint (m) & 3.77 & 2.0 & 2.5 & 2.8 & 3.5 & 2.6 & 2.5 \\
\hline
Aspect ratio $H/D$ & 0.60 & 1.25 & 1.40 & 1.07 & 0.57 & 0.88 & 1.60 \\
\hline
TWR & 2.0 & 1.12 & 1.08 & 2.3 & N/A & N/A & N/A \\
\hline
\end{tabular}
\end{table*}

\section{Propulsion System}
\label{sec:propulsion}

\subsection{YUNT~V0 Engine}
\label{sec:main_engine}

\begin{figure}[t]
\centering
\begin{minipage}[b]{0.48\columnwidth}
\centerline{\includegraphics[height=5.5cm]{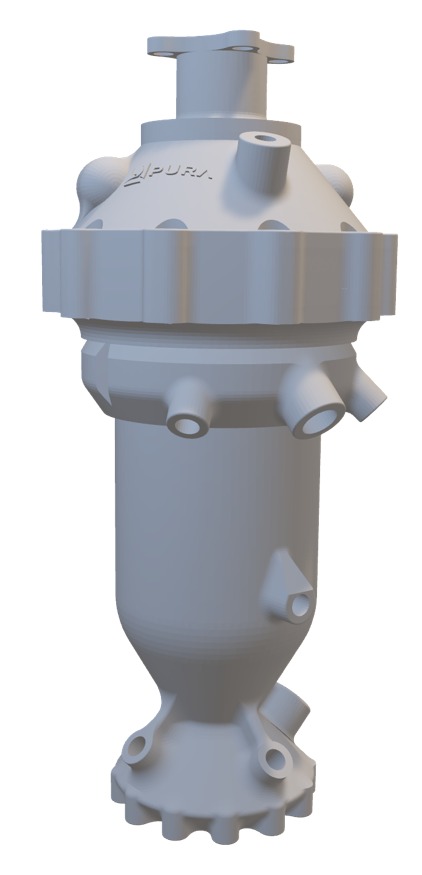}}
\centerline{\small (a)}
\end{minipage}
\hfill
\begin{minipage}[b]{0.48\columnwidth}
\centerline{\includegraphics[height=5.5cm]{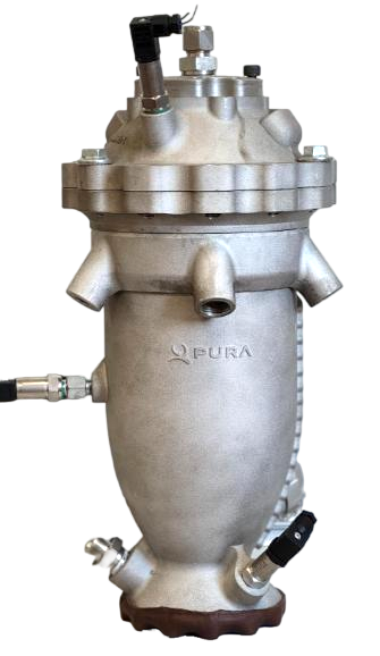}}
\centerline{\small (b)}
\end{minipage}
\caption{YUNT~V0 engine: (a)~CAD model; (b)~printed test article.}
\label{fig:yuntv0}
\end{figure}

The propulsion system centers on the YUNT V0 (Fig.~\ref{fig:yuntv0}), a throttleable, regeneratively cooled liquid rocket engine employing a green storable bipropellant. A pintle-injector baseline was selected for its proven heritage in throttleable engines and suitability for wide operating envelopes~\cite{b_dressler2000, b_casiano2009, b_kang2022}, rather than single-point operation. Green storable propellants eliminate cryogenic storage complexity and legacy hazard concerns, significantly improving ground handling and test readiness~\cite{b_mclean2014}. Chamber thermal management combines regenerative cooling with supplementary film cooling in high-stress regions~\cite{b_chambers1972}. Additive manufacturing enables rapid iteration, reduces component count, and allows integrated cooling passages infeasible with conventional fabrication~\cite{b_gradl2018}. Key performance specifications are summarized in Table~\ref{tab:engine_specs}.

\begin{table}[t]
\centering
\renewcommand{\arraystretch}{1.4}
\caption{YUNT V0 Engine Specifications}
\label{tab:engine_specs}
\begin{tabular}{|c|c|}
\hline
\textbf{Parameter} & \textbf{Value} \\
\hline
Engine type & Bipropellant (green storable) \\
\hline
Max thrust $\rho_{\max}$ & 5200 N \\
\hline
Min thrust $\rho_{\min}$ & 800 N \\
\hline
Throttle ratio & 6.5:1 \\
\hline
Specific impulse $I_{\text{sp}}$ & 210 to 240 s \\
\hline
Dead-zone threshold $T_{\text{dead}}$ & 600 N \\
\hline
Gimbal authority & $\pm7^\circ$ \\
\hline
Thrust error (1$\sigma$) & 1.7\% \\
\hline
Pointing error (1$\sigma$) & 0.5$^\circ$ \\
\hline
\end{tabular}
\end{table}

\begin{figure}[t]
\centerline{\includegraphics[width=\columnwidth]{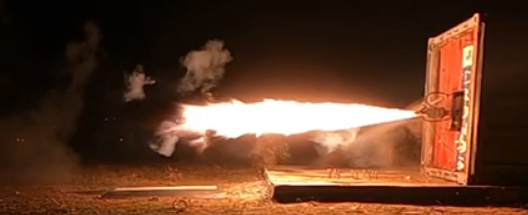}}
\caption{YUNT~V0 hot-fire test.}
\label{fig:hotfire}
\end{figure}

\begin{figure}[t]
\centerline{\includegraphics[width=\columnwidth]{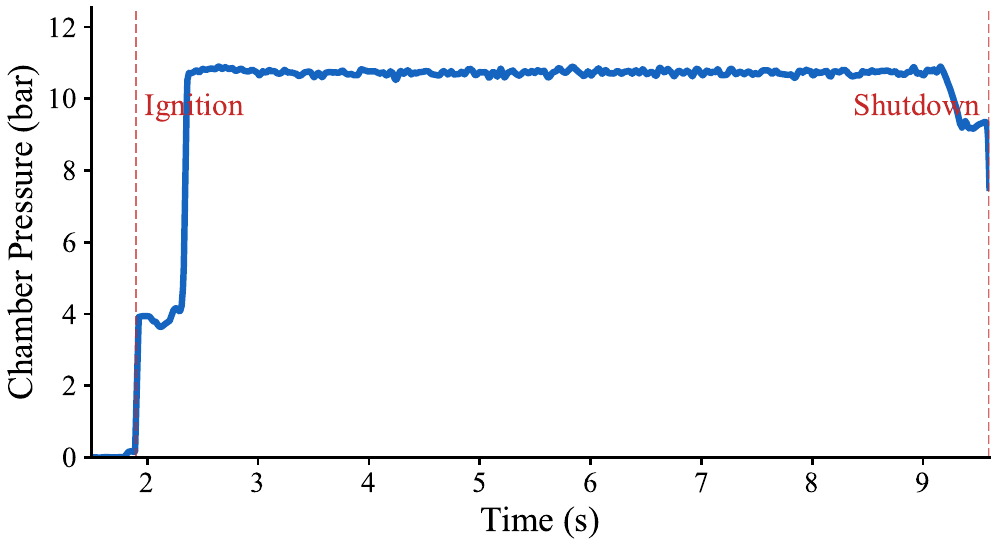}}
\caption{Chamber-pressure time history: ignition, steady-state, and shutdown.}
\label{fig:pc_time}
\end{figure}

Technology maturation followed a staged experimental campaign structured to retire dominant technical risks in sequence: short-duration firings for ignition timing and transient sequencing, mid-duration tests for performance characterization and thermal-model anchoring, and multi-point operation at discrete throttle settings spanning the targeted envelope. Across 16~firings (Fig.~\ref{fig:hotfire}), the engine delivered stable and repeatable operation over a thrust range of $1{,}260$ to $5{,}630$~N at chamber pressures up to 20~bar (exceeding design specifications of 800--5200~N), achieving a peak $c^*$~efficiency of 96.4\%. Dynamic chamber-pressure perturbations remained below 1.5\% of the mean $p_c$ (RMS), as shown in Fig.~\ref{fig:pc_time}, a practical indicator of stable injector-chamber coupling. Time-resolved throttling transients and multi-restart qualification are the focus of ongoing work.

\subsection{Thrust Vector Control and Dead Zone}
\label{sec:tvc_dz}

Thrust pointing is achieved via a two-axis gimbal mechanism providing $\pm 7^\circ$ mechanical authority. Combined with the vehicle's maximum body tilt of $34^\circ$, the effective guidance tilt constraint is $\theta_{\max} = 29^\circ$ after reserving $5^\circ$ for TVC correction authority.

A critical propulsion non-ideality for guidance is the dead zone: for commanded thrust below $T_{\text{dead}} = 600$~N, the actual engine output is unreliable due to injector flow instability and incomplete combustion at low propellant flow rates. Since $T_{\text{dead}}/\rho_{\min} = 75\%$, the dead zone occupies a significant fraction of the throttle range, making careful handling of near-floor commands essential for the guidance algorithm (Section~\ref{sec:gnc}).

\subsection{Propellant System}
\label{sec:feed_system}

The vehicle uses a pressure-fed propellant delivery system with green storable bipropellants. The symmetric tank placement about the vehicle's longitudinal axis minimizes lateral CoM offset during propellant depletion, as confirmed by the near-zero $\text{CoG}_z$ shift in Table~\ref{tab:cg_inertia}. The total propellant load of 85.5~kg provides a theoretical $\Delta V$ of:
\begin{equation}
    \Delta V = I_{\text{sp}} \, g_0 \ln\!\left(\frac{m_0}{m_{\text{dry}}}\right) = 225 \times 9.81 \times \ln\!\left(\frac{259.5}{174}\right) \approx 882~\text{m/s},
\end{equation}
using a nominal $I_{\text{sp}} = 225$~s (midpoint of the operating range). This budget provides substantial margin for multiple ground test flights with vertical ascent, hover, lateral translation, and soft landing profiles.

\section{Guidance, Navigation, and Control}
\label{sec:gnc}

We formulate the powered descent problem as an SOCP solvable in near-real-time, incorporating the propulsion and mass properties from Sections~\ref{sec:design} and~\ref{sec:propulsion}.

\subsection{Dynamics and Constraints}
\label{sec:dynamics}

The translational dynamics of the lander with position $\bm{r}$ and velocity $\bm{v}$ in a landing-site-centered frame are:
\begin{align}
    \dot{\bm{r}}(t) &= \bm{v}(t), \label{eq:rdot} \\
    \dot{\bm{v}}(t) &= \frac{\bm{T}(t)}{m(t)} + \bm{g}, \label{eq:vdot}
\end{align}
where $\bm{T}$ is the thrust vector, $m(t)$ is the vehicle mass, and $\bm{g} = [0, 0, -g_{\text{moon}}]^\top$ with $g_{\text{moon}} = 1.625$~m/s$^2$. Mass depletion follows from the rocket equation:
\begin{equation}
    \dot{m}(t) = -\frac{\|\bm{T}(t)\|_2}{I_{\text{sp}} \, g_0}. \label{eq:mdot}
\end{equation}

The engine thrust magnitude is bounded by $\rho_{\min} \leq \|\bm{T}\|_2 \leq \rho_{\max}$, where the lower bound is the key source of nonconvexity. A tilt constraint $\cos\theta_{\max} \|\bm{T}\|_2 \leq T_3$ and a glide-slope constraint $\|\bm{r}_{1:2}\|_2 \leq r_3 \tan\gamma_{\text{gs}}$ are both second-order cone (SOC) representable and hence convex. The parameters $\rho_{\min}$, $\rho_{\max}$, and $\theta_{\max}$ are set by the propulsion system design; the mass bounds $m_{\text{dry}} \leq m(t) \leq m_0$ are set by the lander configuration.

The minimum-fuel powered descent problem is then
\begin{subequations}
\label{eq:nonconvex}
\begin{align}
\min_{\bm{T}(\cdot),\, t_f} \;\; & -m(t_f) \label{eq:ncvx_obj} \\
\text{s.t.} \;\; & \dot{\bm{r}} = \bm{v}, \quad \dot{\bm{v}} = \bm{T}/m + \bm{g}, \label{eq:ncvx_trans} \\
& \dot{m} = -\|\bm{T}\|_2/(I_{\text{sp}} g_0), \label{eq:ncvx_mass} \\
& \rho_{\min} \leq \|\bm{T}\|_2 \leq \rho_{\max}, \label{eq:ncvx_thrust} \\
& \cos\theta_{\max}\,\|\bm{T}\|_2 \leq T_3, \label{eq:ncvx_tilt} \\
& \|\bm{r}_{1:2}\|_2 \leq r_3 \tan\gamma_{\text{gs}}, \label{eq:ncvx_gs} \\
& m_{\text{dry}} \leq m(t) \leq m_0, \label{eq:ncvx_mbnd} \\
& \bm{r}(0)=\bm{r}_0,\; \bm{v}(0)=\bm{v}_0,\; m(0)=m_0, \label{eq:ncvx_ic} \\
& \bm{r}(t_f)=\bm{r}_f,\; \bm{v}(t_f)=\bm{v}_f. \label{eq:ncvx_tc}
\end{align}
\end{subequations}
Problem~\eqref{eq:nonconvex} is nonconvex due to the lower bound in~\eqref{eq:ncvx_thrust}, the bilinear term $\bm{T}/m$ in~\eqref{eq:ncvx_trans}, the nonlinear $\|\bm{T}\|_2/m$ coupling in~\eqref{eq:ncvx_mass}, and the free final time $t_f$.

\subsection{Convexification Framework}
\label{sec:convexification}

Table~\ref{tab:nonconvex_summary} summarizes the four sources of nonconvexity in~\eqref{eq:nonconvex} and the convexification technique applied to each.

\begin{table}[t]
\centering
\renewcommand{\arraystretch}{1.4}
\caption{Sources of Nonconvexity and Convexification Techniques}
\label{tab:nonconvex_summary}
\begin{tabular}{|c|c|c|}
\hline
\textbf{Source} & \textbf{Type} & \textbf{Technique} \\
\hline
\makecell{Thrust lower bound\\$\|\bm{T}\| \geq \rho_{\min}$} & \makecell{Nonconvex\\constraint} & \makecell{Lossless\\convexification} \\
\hline
\makecell{Thrust-mass\\coupling $\bm{T}/m$} & \makecell{Bilinear\\dynamics} & \makecell{Log-mass\\substitution} \\
\hline
\makecell{Mass depletion\\$\|\bm{T}\|/m$} & \makecell{Nonlinear\\dynamics} & \makecell{Taylor\\linearization} \\
\hline
\makecell{Free final time $t_f$} & \makecell{Variable\\horizon} & \makecell{Time norm.\\+ SCvx} \\
\hline
\makecell{Thruster dead zone} & \makecell{Nonconvex\\set} & \makecell{Strictly-on\\constraint} \\
\hline
\end{tabular}
\end{table}

\begin{figure}[t]
\centerline{\includegraphics[width=\columnwidth]{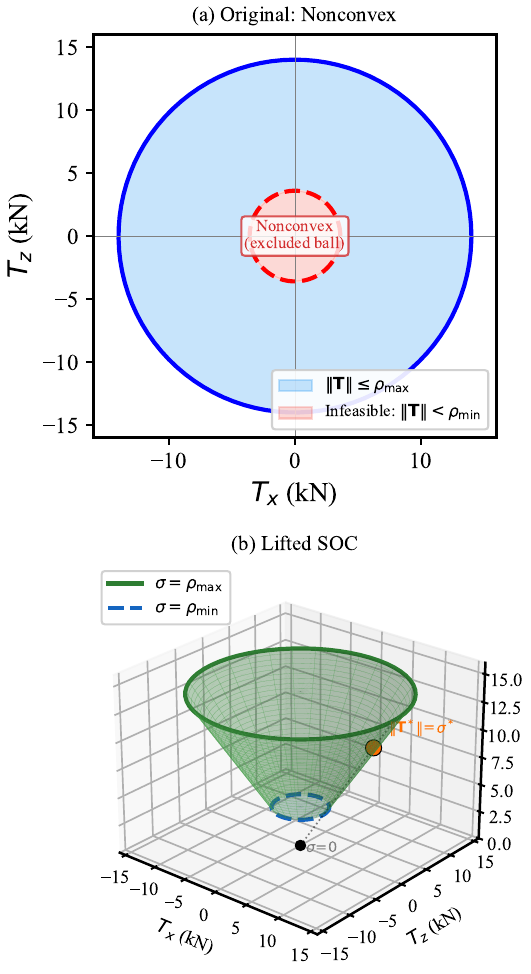}}
\caption{Lossless convexification: (a) nonconvex set; (b) lifted SOC.}
\label{fig:thrust_constraint}
\end{figure}

Lossless convexification~\cite{b_acik2007, b_acik2011} replaces the nonconvex thrust lower bound via a slack $\sigma$: $\|\bm{T}\|_2 \leq \sigma$, $\rho_{\min} \leq \sigma \leq \rho_{\max}$, with $\dot{m} = -\sigma/(I_{\text{sp}} g_0)$. The log-mass substitution $z = \ln m$, Taylor linearization of $e^{-z}$, and time normalization $\tau \in [0,1]$ yield an SOCP at each SCvx iteration $k$ with decision variables $\{\bm{r}_j, \bm{v}_j, \bm{u}_j, z_j, \sigma_j, t_f\}$ and objective $\min\,{-z_N} + w_\nu\sum_j\|\bm{\nu}_j\|_1$. The strictly-on bound $\rho_{\min}\hat{\mu}_j^{(k)} \leq \sigma_j \leq \rho_{\max}\hat{\mu}_j^{(k)}$ keeps thrust above the 600~N dead zone since $\rho_{\min} = 800$~N $> T_{\text{dead}}$. Trust regions ensure convergence in 5--15 iterations~\cite{b_mao2016}, each solved in milliseconds by Clarabel~\cite{b_clarabel} via CVXPY~\cite{b_cvxpy}. Fig.~\ref{fig:thrust_constraint} illustrates the geometric lifting.

\section{Simulation and Results}
\label{sec:results}

\subsection{Simulation Setup}

Simulations use BUG/YUNT V0 parameters (Tables~\ref{tab:bug_summary}, \ref{tab:engine_specs}). The truth model propagates 3-DoF dynamics via RK4 ($\Delta t = 0.1$~s); perturbations are detailed in Table~\ref{tab:mc_perturbations}. The SCvx solver converges in 8--12 iterations (residuals $<\!10^{-10}$) with $\sim$200~ms solve time per trajectory.

\subsection{Tilt-Angle Design Envelope}

At maximum allowable tilt $\theta_{\max}$ combined with minimum throttle $\rho_{\min}$, the vertical component $\rho_{\min}\cos\theta_{\max}$ may equal or exceed vehicle weight, eliminating net downward acceleration and mandating coast arcs. The critical feasibility boundary is:
\begin{equation}
    \theta_{\max}^{\text{crit}} = \arccos\!\left(\frac{m \, g_{\text{moon}}}{\rho_{\min}}\right) \label{eq:feasibility}
\end{equation}
For the BUG vehicle ($\rho_{\min} = 800$ N, $m_0 = 259.5$ kg), $\theta_{\max}^{\text{crit}} \approx 58^\circ$. The TVC configuration ($\theta_{\max} = 60^\circ$) exceeds this boundary, enabling continuous-thrust descent; ACS ($\theta_{\max} = 29^\circ$) operates below it, requiring burn-coast-burn profiles.

\begin{figure}[t]
\centerline{\includegraphics[width=\columnwidth]{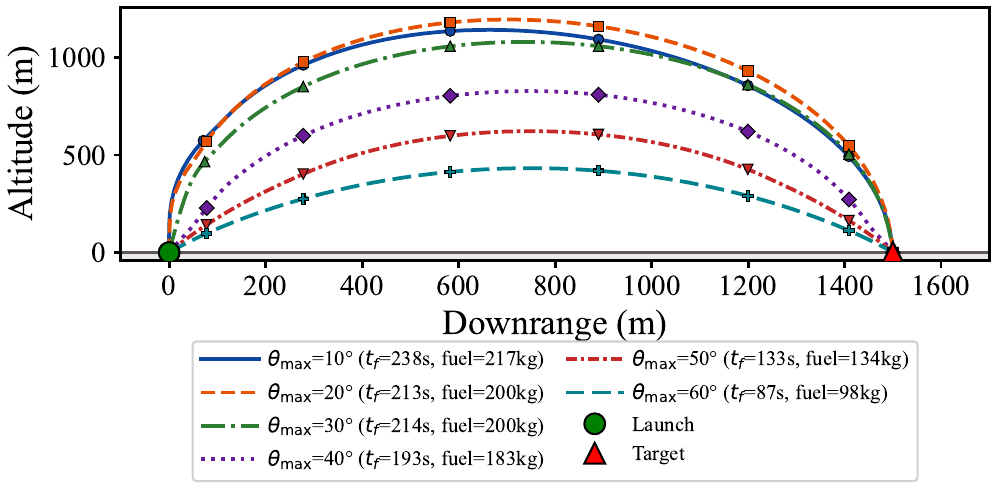}}
\caption{Hop trajectories for $\theta_{\max} \in \{10^\circ,\dots,60^\circ\}$. Fuel cost and flight time decrease as tilt authority increases.}
\label{fig:theta_combined}
\end{figure}

Fig.~\ref{fig:theta_combined} confirms the sharp feasibility transition across the sweep, validating that $\theta_{\max}^{\text{crit}}$ is the design-driving boundary: $\rho_{\min}$ directly determines the minimum tilt authority required for continuous-thrust descent.

\subsection{Monte Carlo Robustness Validation}

$N = 1000$ closed-loop realizations are run with the perturbation model in Table~\ref{tab:mc_perturbations} and ten rare failure modes (engine shutoff, thrust degradation, power brownout, sensor failure, RCS stuck, propellant leak, gimbal seizure, computer reset, slosh, nozzle erosion; $\sim$2.5\% total activation per run). SCvx re-solves every 10~s from the perturbed state.

\begin{table}[t]
\centering
\renewcommand{\arraystretch}{1.4}
\caption{Monte Carlo Perturbation Model}
\label{tab:mc_perturbations}
\begin{tabular}{|c|c|}
\hline
\textbf{Perturbation} & \textbf{Value} \\
\hline
Initial position dispersions & $\sigma_r = 10$ m/axis \\
\hline
Initial velocity dispersions & $\sigma_v = 2$ m/s/axis \\
\hline
Initial mass uncertainty & $\sigma_m = 1\%$ \\
\hline
Thrust magnitude bias & 3\% (per-run) \\
\hline
Thrust magnitude noise & 1\% (per-step) \\
\hline
Thrust pointing bias & $1^\circ$ \\
\hline
Thrust pointing jitter & $0.5^\circ$ (per-step) \\
\hline
Gravity model error & 0.5\% \\
\hline
Specific impulse uncertainty & 2\% \\
\hline
Process noise (accel) & $0.05$ m/s$^2$ \\
\hline
\end{tabular}
\end{table}

\begin{figure}[t]
\centerline{\includegraphics[width=\columnwidth]{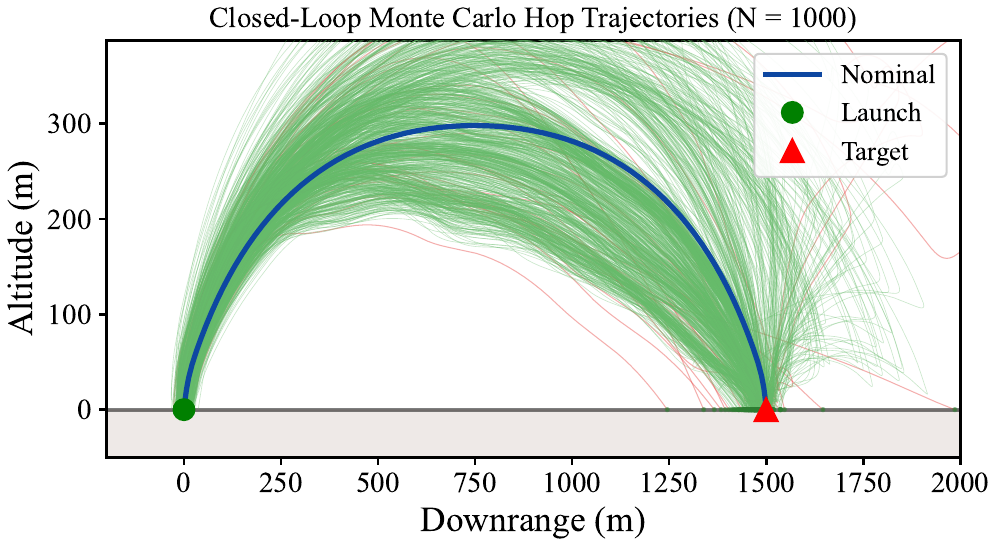}}
\caption{Closed-loop Monte Carlo (N=1000): nominal (green), failure modes (orange), reference (blue). CL replanning: $<$12~m vs.\ 147~m open-loop.}
\label{fig:mc_cl_trajectories}
\end{figure}

Closed-loop SCvx replanning (200~ms solve, 10~s cadence) achieves $<$12~m mean landing error vs.\ 147~m open-loop, correcting accumulated drift during descent.

\section{Conclusion}
\label{sec:conclusion}

This paper presented an integrated lander-propulsion-GNC framework for autonomous lunar powered descent on the BUG vehicle and YUNT~V0 engine. The SCvx algorithm addresses all nonconvexities with a strictly-on constraint maintaining thrust above 600~N throughout descent. Parametric analysis revealed a tilt-angle feasibility boundary coupling throttle ratio to pointing authority. Monte Carlo validation confirmed onboard replanning reduces landing dispersion by over an order of magnitude, achieving sub-50-meter precision. This collaborative effort across Puura Inc., Turkuzaysan Inc., Final Proximity Inc., and TUBITAK demonstrates how integrated vehicle design, propulsion, and guidance advances autonomous lunar landing. Future work targets full 6-DoF SCvx validation through ground-based VTVL test campaigns.

\section*{Declaration of AI-Assisted Writing}

The authors used Claude (Anthropic) for clarity and organization. All scientific work is original. Human judgment controlled technical content.

\balance

\end{document}